# A C-band microwave rectenna using aperture-coupled antenna array and novel Class-F rectifier with cavity

Chengyang Yu, Feifei Tan and Changjun Liu*

*School of Electronics and Information Engineering, Sichuan University, Chengdu 610064, China*

A rectenna (rectifying antenna) is usually applied to a microwave power transmission (MPT) system as a terminal, which receives microwave (MW) power and converts them into DC power. A 5.8 GHz aperture-coupled patch antenna array is developed with a gain of 19.5 dBi, which is a part of the rectenna. A novel Class-F rectifier with a series diode is proposed to achieve a high rectifying efficiency based on harmonic termination techniques. The input circuit of the proposed rectifier is taken into consideration to reach the Class-F condition, while mostly only the output circuit is involved in conventional Class-F designs. A low-pass filter based on defected ground structure is implemented, which not only blocks the harmonics produced in rectifying but also presents desired impedance at harmonics. The current and voltage waveforms on the diode are well enforced. The radiation feature of the rectifier is presented as well. It shows that an optimized cavity, which shields the parasite radiation, may achieve a 5% enhancement on the MW-to-DC conversion efficiency. A C-band rectifier based on HSMS-286 Schottky diode is fabricated and measured. A MPT simple demo system is constructed with the proposed rectifier and antenna. A measured maximum MW-to-DC conversion efficiency of 72% is achieved on the rectifier. The design method is extremely valuable in a high conversion efficiency rectenna design based on harmonic termination techniques.

**Keywords:** rectenna; series diode; Class-F; cavity; microwave power combining

## 1. Introduction

Recently rectifying antenna (rectenna) has become a very active area of research because of its numerous applications, such as space solar power satellite, RFID tags, wireless sensor, energy harvesting, and other microwave power transmission (MPT) systems.[1–3] A microwave rectenna is a receiving antenna associating a rectifier which converts microwave (MW) power into DC power.[4] It has been widely studied at 915 MHz [5] and 2.45 GHz.[6] However, the rectenna physical dimensions at these bands are a little large for large-scale applications. Therefore, the industry–science–medical (ISM) band of 5.8 GHz [7] is a suitable candidate due to the reduction of the antenna size and lower cost compared with other higher ISM bands.[8]

In order to achieve a good transmission link between transceiver devices, a high-gain directional antenna and a rectifier with high MW-to-DC conversion efficiency are necessary. Several antenna types have been applied to MPT systems such as dipole antennas,[9] loop antennas,[10] and slot antennas.[11] In this paper, a patch antenna array [12, 13] is utilized to collect more power for a far-field transmission with lower incident density and an aperture-coupled feed is employed to reduce the mutual interference between the antenna and feed circuit.

*Corresponding author. Email: cjliu@ieee.org

For a power transmission system, the conversion efficiency between different types of energy is an essential factor. Different from a bridge rectifier in low-frequency analog circuit, single-shunt rectifiers are often applied due to its low loss among the existing MW rectifier studies. The dissipation model of rectifying diode and harmonic recycling method are utilized to analyze enhancement on rectifying efficiency. As presented in Ref. [7], a normal single-shunt rectifier is composed of an input filter, a shunt diode, and an output filter connected to a DC load. Composed of a shunt capacitor and a quarter wavelength transmission line, the output DC filter shows an impedance of zero at even harmonics and that of infinity at odd harmonics. Thus, the V–I characteristic of the rectifier is very close to a Class-F power amplifier.[14] Recently, Class-F load is developed as the output filter of rectifier by Kyoto University.[15] Class-F rectifier is one kind of resonant rectifiers which contain a series of resonant cells and provides improved efficiency by shaping the voltage and current curves. Resonant rectifiers were originally proposed in [16] in 1987, and Class-E rectifiers were then analyzed and applied in DC–DC converters.[17,18] In a Class-E rectifier, all harmonics are presented in both voltage and current waveforms, but these two waveforms are in-phase quadrature and no power is produced at the harmonic frequencies.

Most designs of conventional Class-F and other types of harmonic loads only focus on the output circuit of a rectifier. In this paper, the input and output circuits are analyzed simultaneously to realize a novel Class-F load for a C-band rectifier with a series diode configuration. For the input termination, a low-pass filter based on defected ground structure (DGS) is implemented, which not only blocks the harmonics produced by diode but also presents desired impedance at harmonics. A full circuit optimization is performed to the rectifier with the consideration of diode package effects. Furthermore, parasitic radiation patterns of the MW rectifier are presented as well. Recycling of the radiation power is important to maximize the conversion efficiency. A cavity is then introduced to the proposed rectifier. About 5% efficiency enhancement is finally obtained by adjusting the cavity. In the end, a MPT demo system is constructed with the proposed rectenna and a maximum MW-to-DC conversion efficiency of 72% is achieved on the proposed rectifier.

## 2. Rectenna design

### *2.1. Antenna design and fabrication*

An aperture-coupled antenna array, which is selected to efficiently receive microwave power at a low power density, is illustrated in Figure 1. It consists of three layers, e.g. the patch layer, feeding layer, and the slot layer used as the ground plane. An aperture-coupled feed is employed for reducing the mutual interference between the antenna and the feed circuit. This structure will also serve to simplify the design processes since the antenna and the rectifying circuit can be designed independently due to the very limited coupling effects. All the layers are printed on two F4B substrates with a dielectric constant of 2.65 and thickness of 1 mm, respectively. Patches are utilized because of its simplicity, low-cost fabrication, and easy integration with microstrip rectifiers.[19] Approximately, the length of the patch can be expressed as

$$L \approx \frac{1}{2}\lambda_g \tag{1}$$

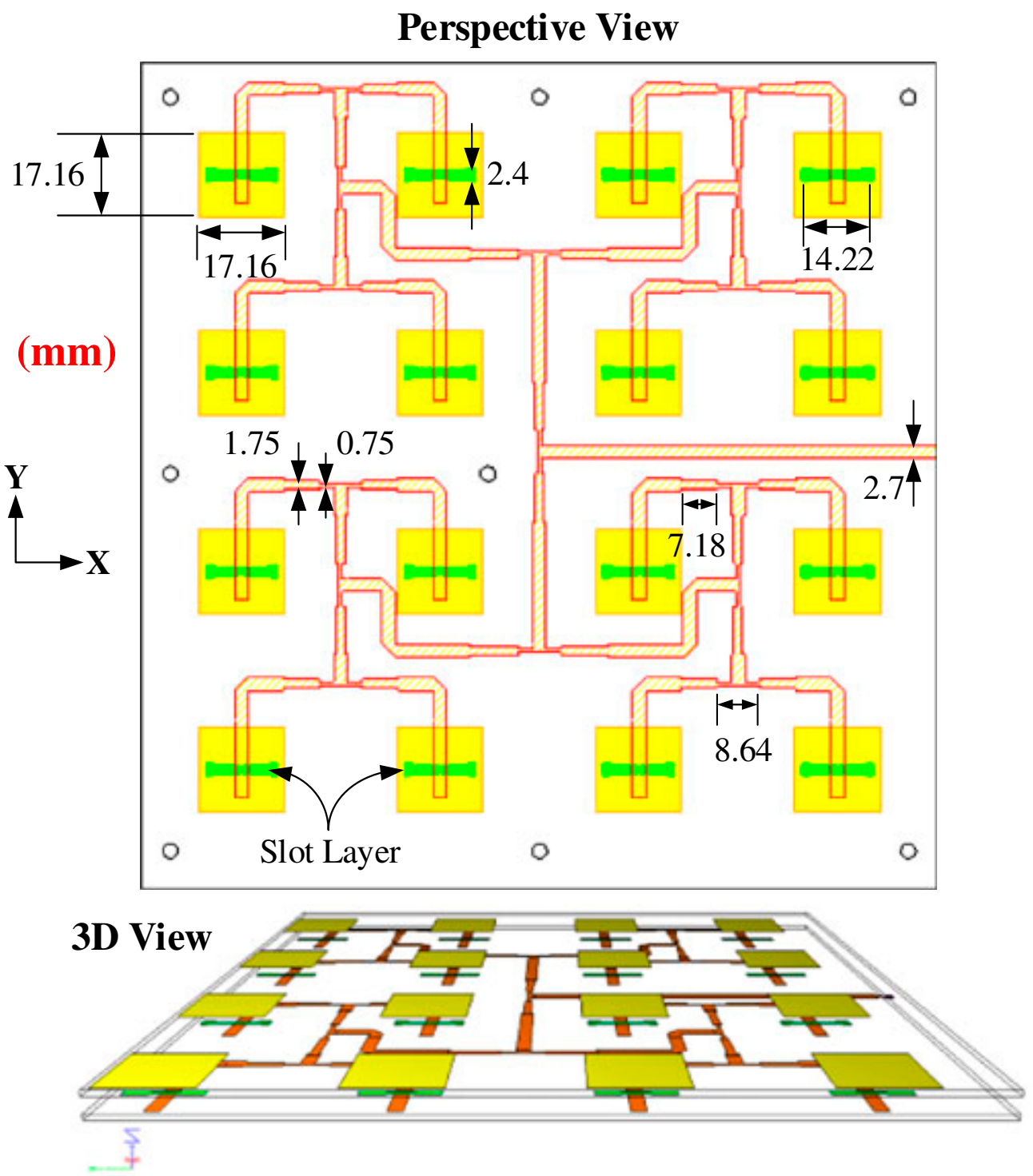

Figure 1. Structure of aperture-coupled antenna array.

The fabricated antenna element is with dimensions of 50 mm by 60 mm and its return loss is shown in Figure 2. The antenna element works well at the operation frequency at 5.8 GHz. Its *E*-plane and *H*-plane normalized radiation patterns are shown in Figure 3. The measured and simulated results agree well and prove the directional characteristic.

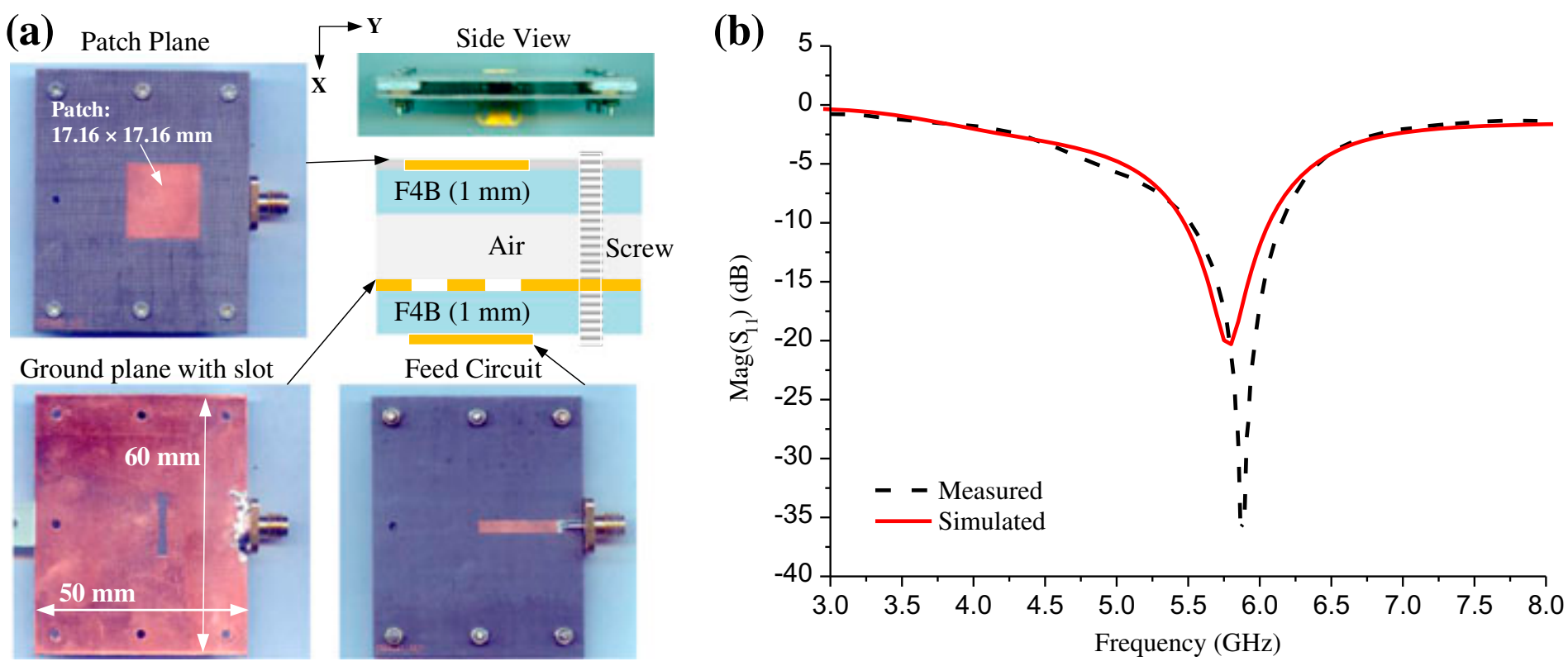

Figure 2. The antenna element: (a) fabrication and (b) return loss.

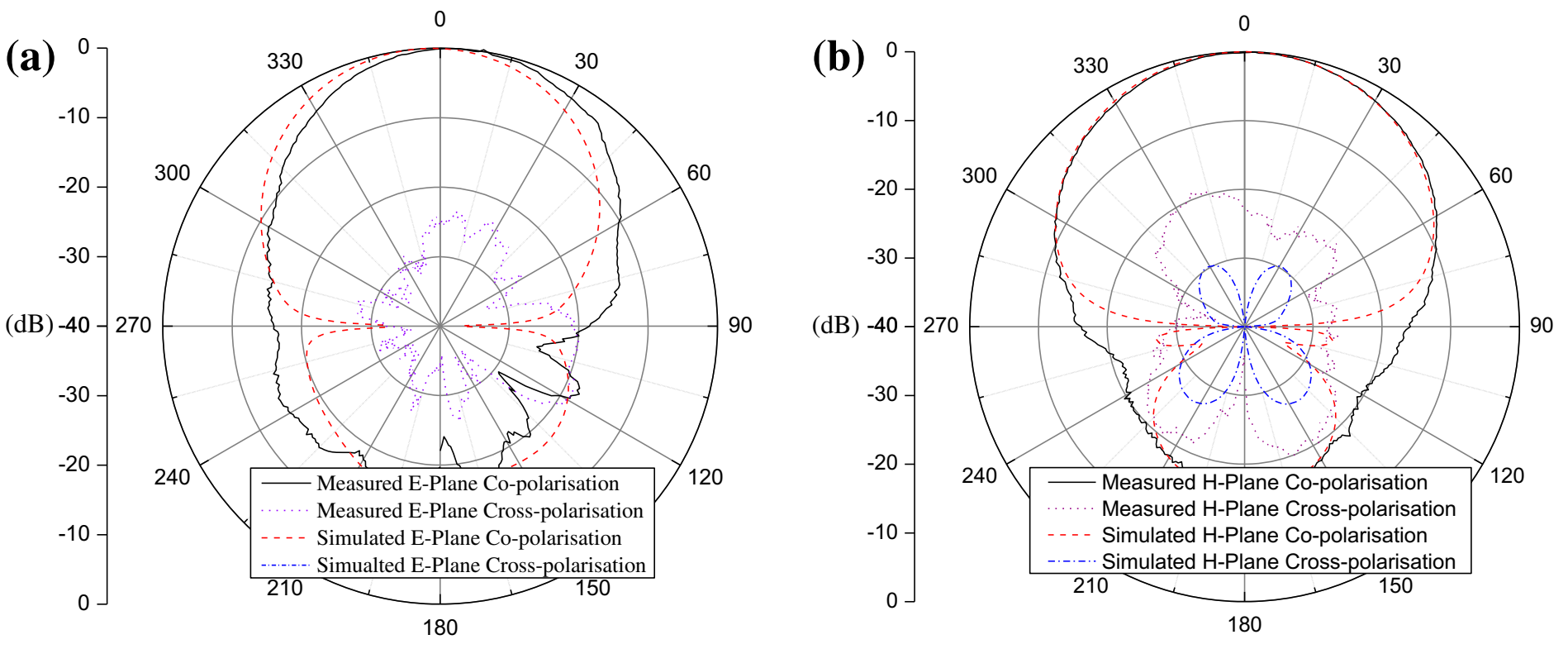

Figure 3. Normalized radiation pattern of the proposed antenna element: (a) *E*-plane and (b) *H*-plane.

For the feed circuits design, the patch element with a 50 Ω input resistance at the edge is connected to the feed line with the characteristic impedance of 50 Ω and transformed through a 70.7 Ω quarter-wavelength transformer to 100 Ω. The 100 Ω lines from two neighboring elements are joined at a T-junction and converted back to a single 50 Ω line. In the next step, neighboring pairs of 50 Ω lines are again transformed through a 70.7 Ω quarter-wavelength transformer and similarly joined at the next T-junction. The process is continued until the final pair of the feed line is joined at the last T-junction. In this case, all of the quarter-wavelength transformers are 70.7 Ω, theoretically. Patch elements and feed circuits are analyzed and optimized using Mentor Graphics IE3D.

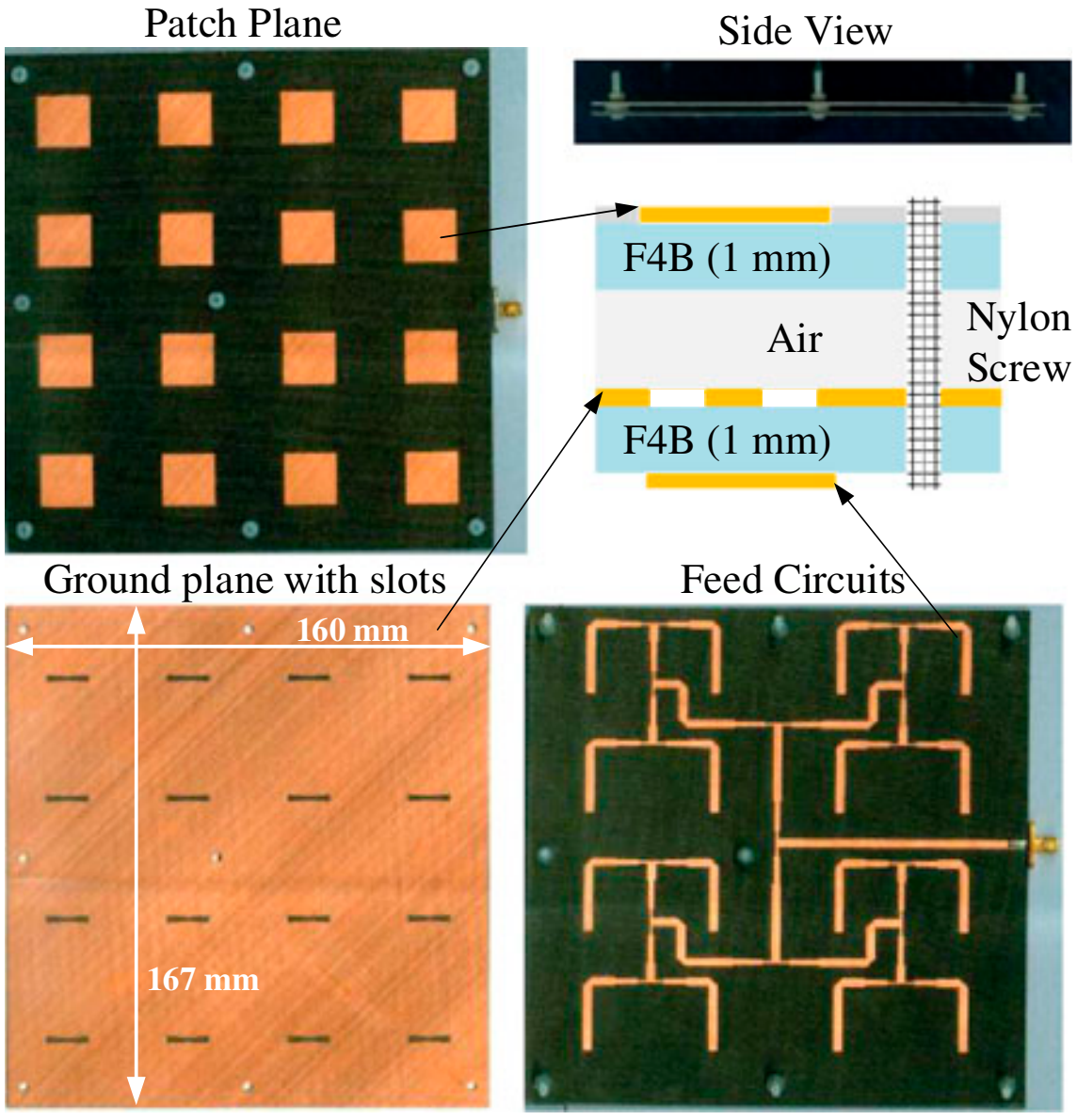

Figure 4. The fabricated antenna array.

The fabricated antenna array is shown in Figure 4. The two substrates are separated by air with distance of 2.7 mm. Nylon screws are used to prop them up. For a large-scale application, as shown in Figure 5, the effect of machining error on return loss is discussed in this paper. This thickness of air layer has an allowable error of at least plus or minus 0.5 mm for the antenna array.

The measured radiation pattern of the proposed antenna can be seen in Figure 6. This antenna array also has a high measured gain of 19.5 dBi at 5.8 GHz.

### 2.2. Rectifier design and fabrication

#### *2.2.1. Novel Class-F rectifier*

Harmonic termination techniques are widely used in power amplifier design, where the current and voltage waveforms are shaped to minimize harmonic loss. A novel Class-F load is constructed to the proposed rectenna based on harmonic termination techniques. In order to obtain a good electrostatic protection for field experiment, a series diode configuration, with a short-ended stub in the front of it, is employed for rectenna design. The input and output networks of diode shown in Figure 7, respectively, represent two ideal terminations at fundamental frequency and its harmonics, which permit voltages to exist at fundamental frequency and odd harmonics. Ideal impedance conditions of the input termination $Z_{d,\mathrm{in}}$ and output termination $Z_{d,\mathrm{out}}$ are expressed by:

$$Z_{d,in}(f) = \begin{cases} Z_{DS}, & f = f_0 \\ 0, & f = 2f_0 \\ \infty, & f = 3f_0 \end{cases} \tag{2}$$

$$Z_{d,out}(f) = \begin{cases} Z_{DL}, & f = f_0 \\ 0, & f = 2f_0 \\ \infty, & f = 3f_0 \end{cases} \tag{3}$$

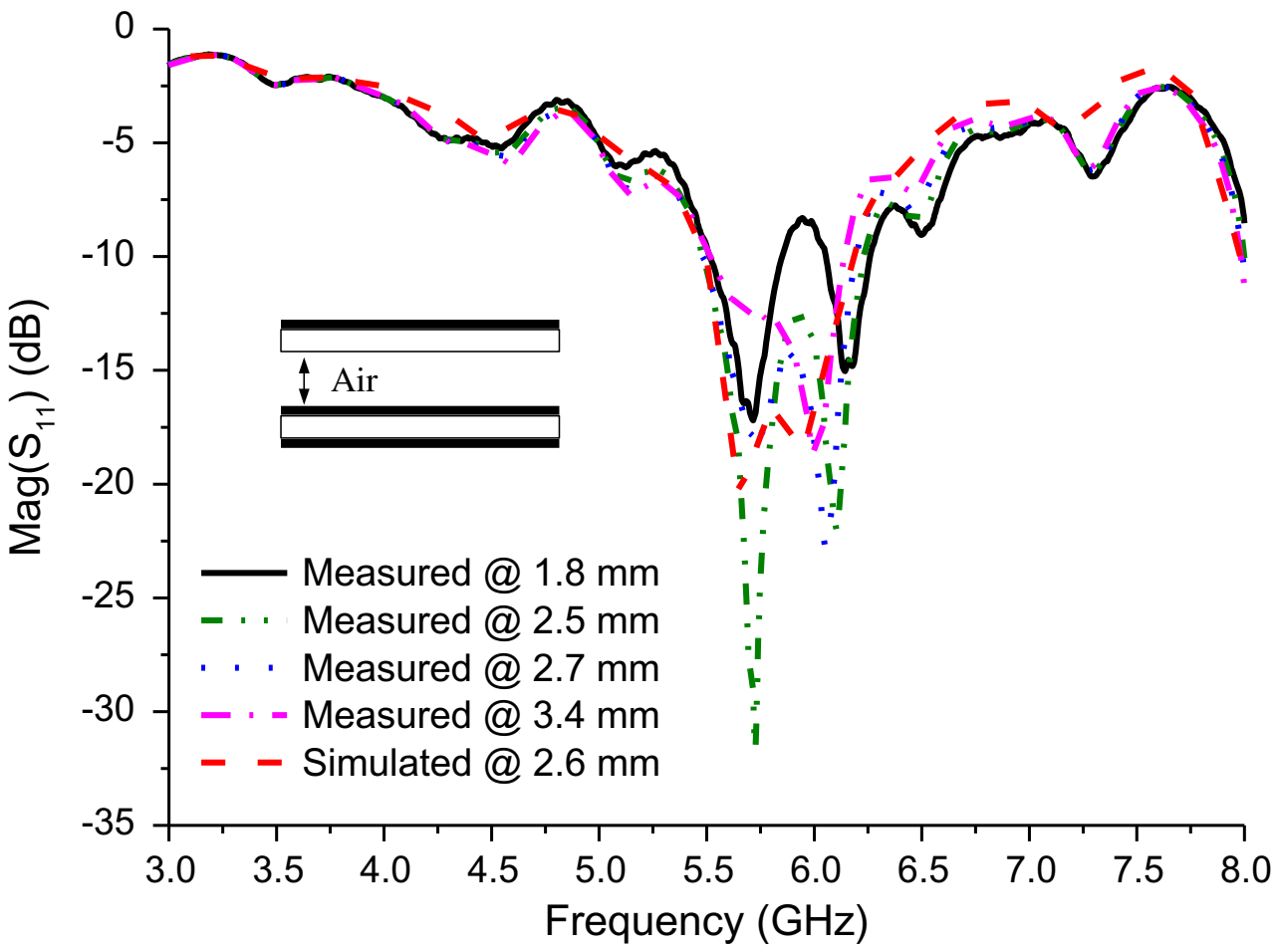

Figure 5. Return loss comparison of the antenna array with different thicknesses of air layer.

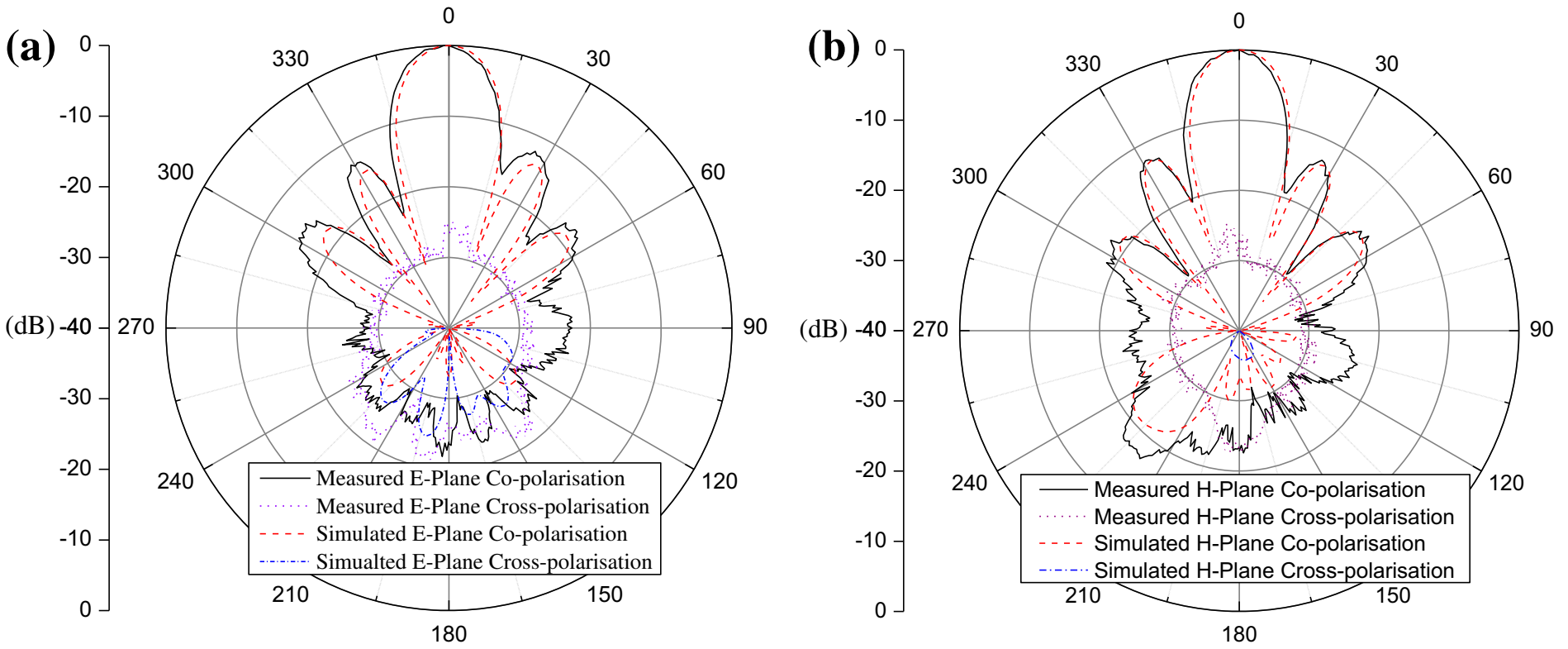

Figure 6. Normalized radiation pattern of the proposed antenna array: (a) *E*-plane and (b) *H*-plane.

In other words, for realizing the Class-F condition, $Z_{d,\mathrm{in}}$ and $Z_{d,\mathrm{out}}$ should act to short circuit at even harmonics other than the fundamental frequency and odd harmonics. If $P_1$ and $P_2$ all act to a short-circuited state, the two quarter-wavelength transformers at the fundamental frequency will provide opposite impedance conversions at odd and even harmonics, respectively. In our work, only 2nd and 3rd harmonics are concerned. Input filter, output filter, and the short-ended stub will provide the needed short circuit for harmonics.

### 2.2.2. *Input termination*

In this paper, the input termination is composed of a source resistance, an input filter, a short-ended stub, and a series microstrip line connected to one pin of the series diode. The short-ended stub and series microstrip line are typically a quarter-wavelength at the fundamental frequency. The ideal impedance at point $P_1$ is zero at 2nd harmonic. Thus, the input filter should perform the following three tasks: (1) impedance matching, (2) pass the power at fundamental frequency, and (3) also prefer a short circuit at 3rd harmonic. This filter design is based on DGS [20].

Split ring resonators are utilized to pass the power at 5.8 GHz and especially provide a short circuit for 3rd harmonic. Modeling and optimization processes are performed in IE3D software. The optimized DGS filter is illustrated in Figures 8 and 9.

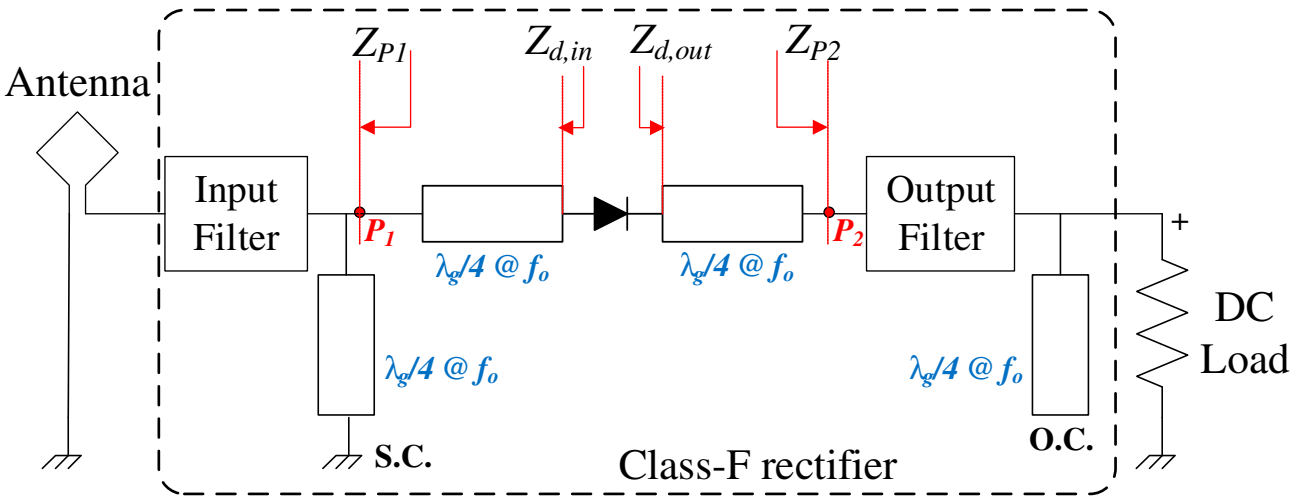

Figure 7. Diagram of Class-F rectenna including input and output terminations.

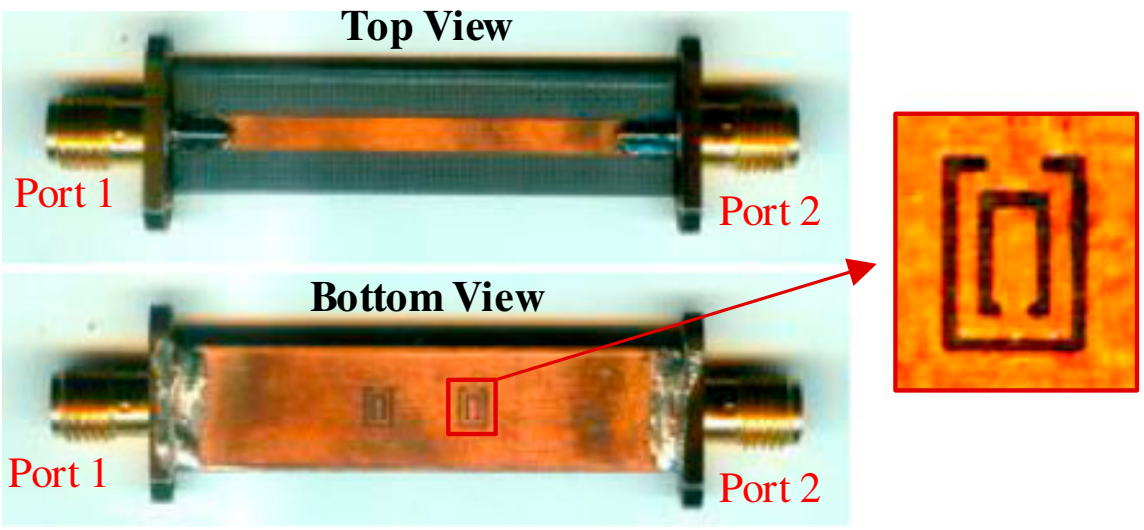

Figure 8. The fabricated DGS filter.

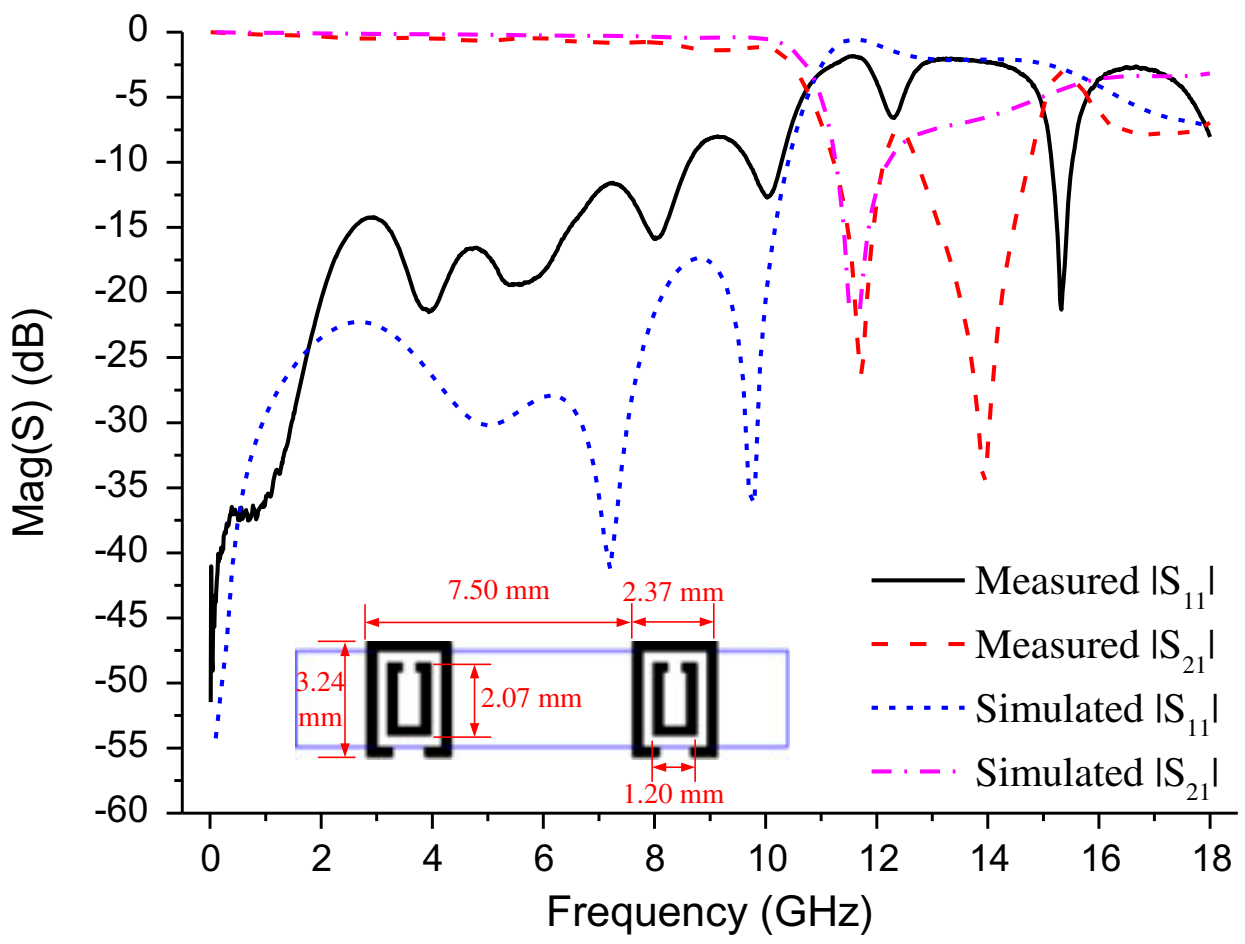

Figure 9. S-parameters of the DGS filter.

The insertion loss from Port 1 to Port 2 of this filter is measured to be 0.46 dB at 5.8 GHz. Meanwhile, as shown in Figure 10, the measured input impedance from Port 2 of this filter is 237 and 17 Ω at the 2nd harmonic and 3rd harmonic, respectively. This indicates, with regard to Equation (2), that the input termination of diode acts to open circuit at 3rd harmonic because of the DGS filter and its associated series microstrip line.

### *2.2.3. Output termination*

The output termination is composed of a DC load, an output filter, an open-ended stub, and a series microstrip line connected to another pin of the series diode. The series microstrip line and open-ended stub are theoretically a quarter-wavelength at the fundamental frequency. The input impedance seen from point $P_2$ should act to short circuit at all frequencies other than the DC signal. Two Murata chip capacitors (GRM1885C1H680JA01) are used as output filter, which avoid microwave power dissipation on the DC load. This also indicates, with regard to Equation (3), that the output termination of diode acts to short circuit at 2nd harmonic and open circuit at 3rd harmonic.

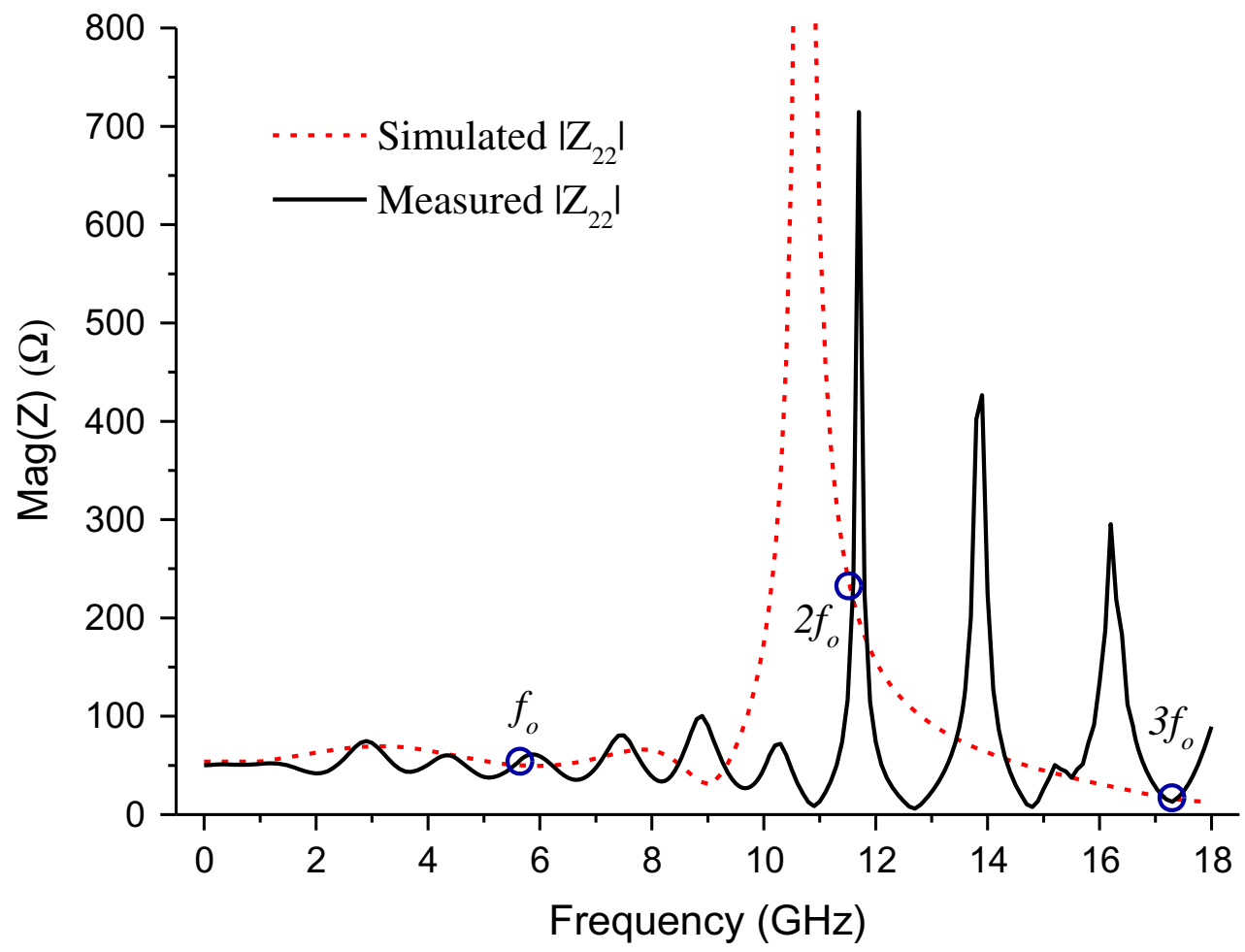

Figure 10. Impedance of the DGS filter.

### 2.2.4. *Rectifying diode*

The rectifying diode is the core element of a rectifier. Schottky diodes are most suitable to microwave rectifiers due to their low turn-on voltage and low junction capacitance. An Avago Schottky diode HSMS-286 is applied to this work with considering on its zero-bias junction capacitance $C_{J0} = 0.18$ pF and series resistance $R_S = 6\ \Omega$. In addition, this series diode is economical for large-scale application. Its other equivalent parameters are shown as follows: the reverse breakdown voltage $V_B = 7$ V, the forward bias turn-on voltage $V_{BI} = 0.65$ V, and the package type SOT-323. This diode model may be found in advanced design system (ADS) component library.

Terminations of Class-F rectifier are discussed above without considering the parasitic effects on rectifying diode. In Ref. [15], separate optimization of Class-F load works well with MA4E1317 diodes, since the package effects of the diodes may be neglected. However, the HSMS-286C diode, used in this paper, usually has a stronger package effects. Equivalent model of the diode package is shown in Figure 11, where $C_P$ is the package capacitance and $L_B$ is the bond-wire inductance, $L_L$, $C_L$ are the lead-frame inductance and capacitance of SOT-323, respectively. The resulting operation state of the diode die does not match to the Class-F load, which will violate the design principle and affect the rectifying efficiency. Thus, a whole circuit optimization operated in Agilent ADS software is preferred. Figure 12 shows the optimized layout of the proposed Class-F rectifier. Locations of lumped capacitors can be adjusted manually to get better measured results.

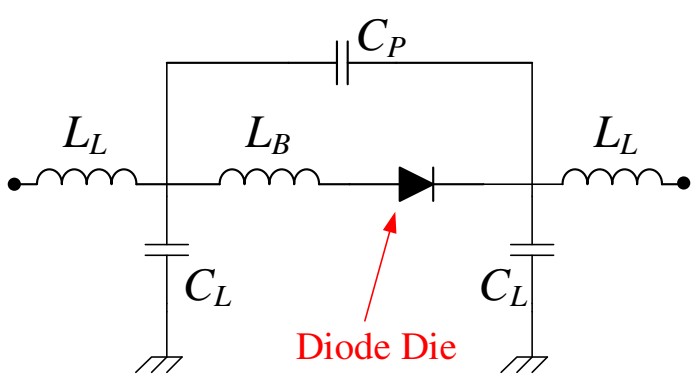

Figure 11. Equivalent model of SOT-323 package.

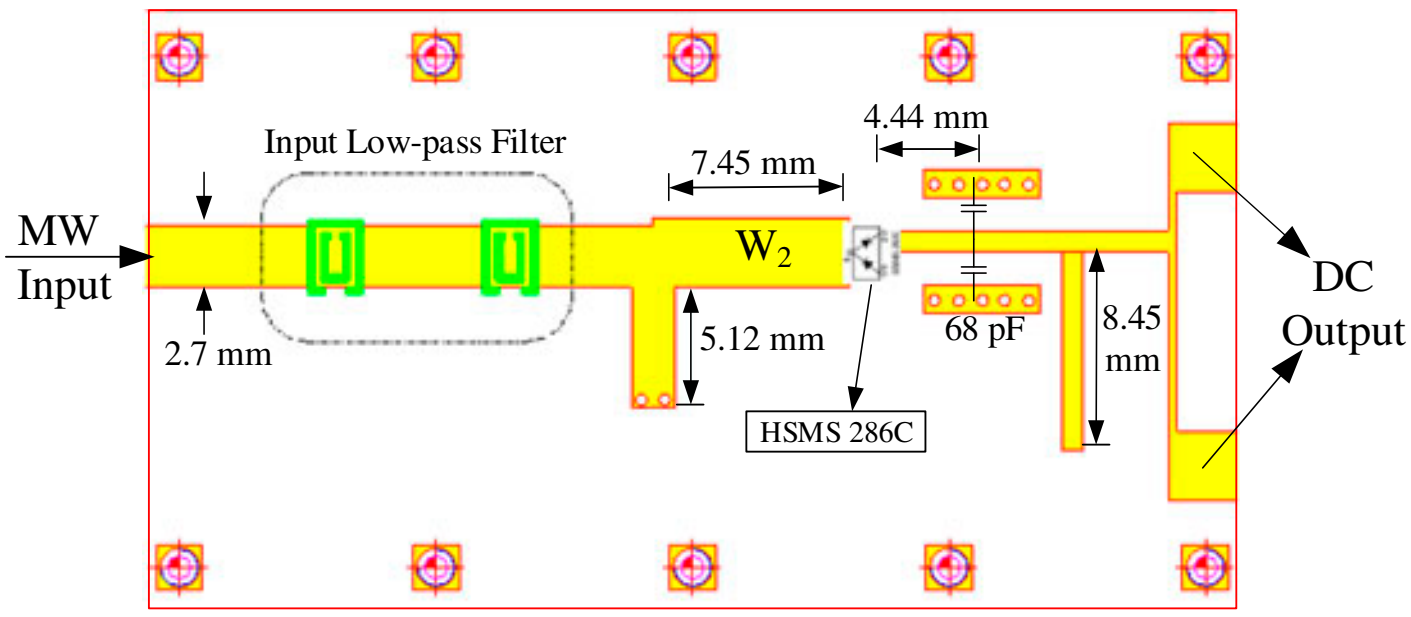

Figure 12. Layout of the proposed Class-F rectifier.

### *2.2.5. Radiation effect and cavity design*

Although harmonic termination techniques are used to enhance rectifying efficiency, microwave power will also radiate into free space with some resonant structures. Figure 13 shows simulated radiation patterns at fundamental frequency and its harmonics. The parallel stubs are predicted to the main radiation source of the proposed rectifier.

In order to reduce power radiations of rectifier, cavity structure is investigated in this paper. The radiation power will be reflected by top cover and trapped in the cavity. Figure 14 shows the 3D model of the rectifier with cavity built in layout of ADS. Cavity walls are consisted of perfect conductor covers and PEC background.

The distance from circuit plane to the cover of cavity is crucial to reflect radiation power. The preferred distance $H_1$ is approximately given by:

$$H_1 \approx \frac{1}{4}\lambda_0 \times \cos\theta \tag{4}$$

where $\theta$ is the angle between $Z$-axis and the maximum radiation direction at fundamental frequency. When properly designed, the conversion efficiency of rectifier will be further enhanced.

The fabricated 5.8 GHz rectifier with cavity is shown in Figure 15. The circuit is also realized on a F4B substrate with a dielectric constant of 2.65 and thickness of 1 mm. A milled groove with a depth of 3.5 mm is constructed on the cavity to match the DGS filter. Figure 16 shows the measurement comparison of the rectifier only and rectifier with cavity. A maximum efficiency of 67.4% is obtained for the rectifier without cavity when the input power is 16 dBm and the DC load is 324 Ω. However, for the rectifier with cavity, a maximum conversion efficiency of 72.1% is achieved at an input power of 15 dBm and a DC load of 380 Ω. In addition, MW-to-DC efficiency variation relative to the air thickness $H_1$ is investigated. As shown in Figure 17, the simulated and measured data match well. These prove that rectifier with cavity can achieve a higher efficiency regardless of radiation. In addition, for a field experiment, environment and electromagnetic interferences can also be effectively prevented using a cavity structure.

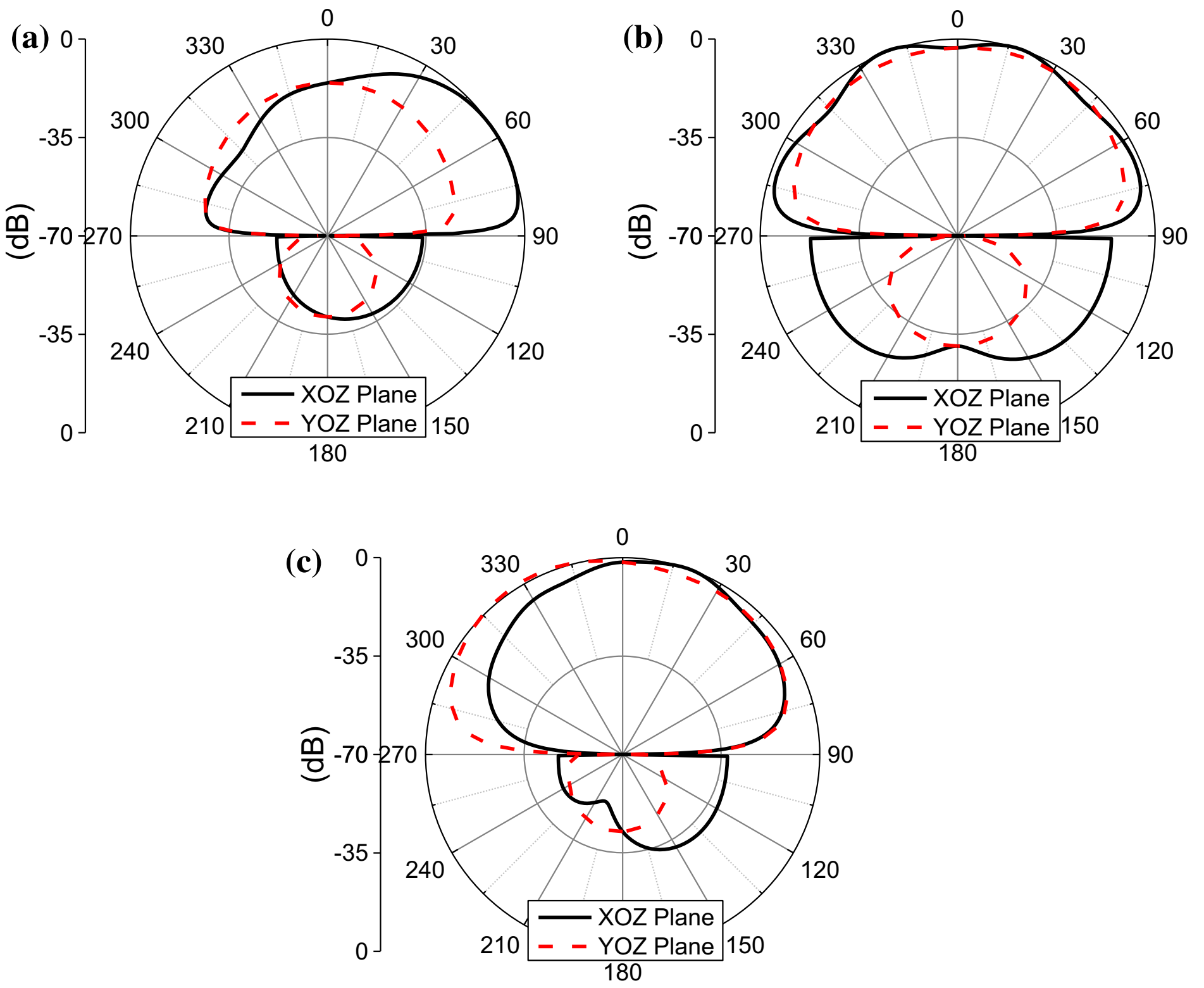

Figure 13. Simulated radiation pattern of the 5.8 GHz rectifier: (a) fundamental frequency, (b) 2nd harmonic, and (c) 3rd harmonic.

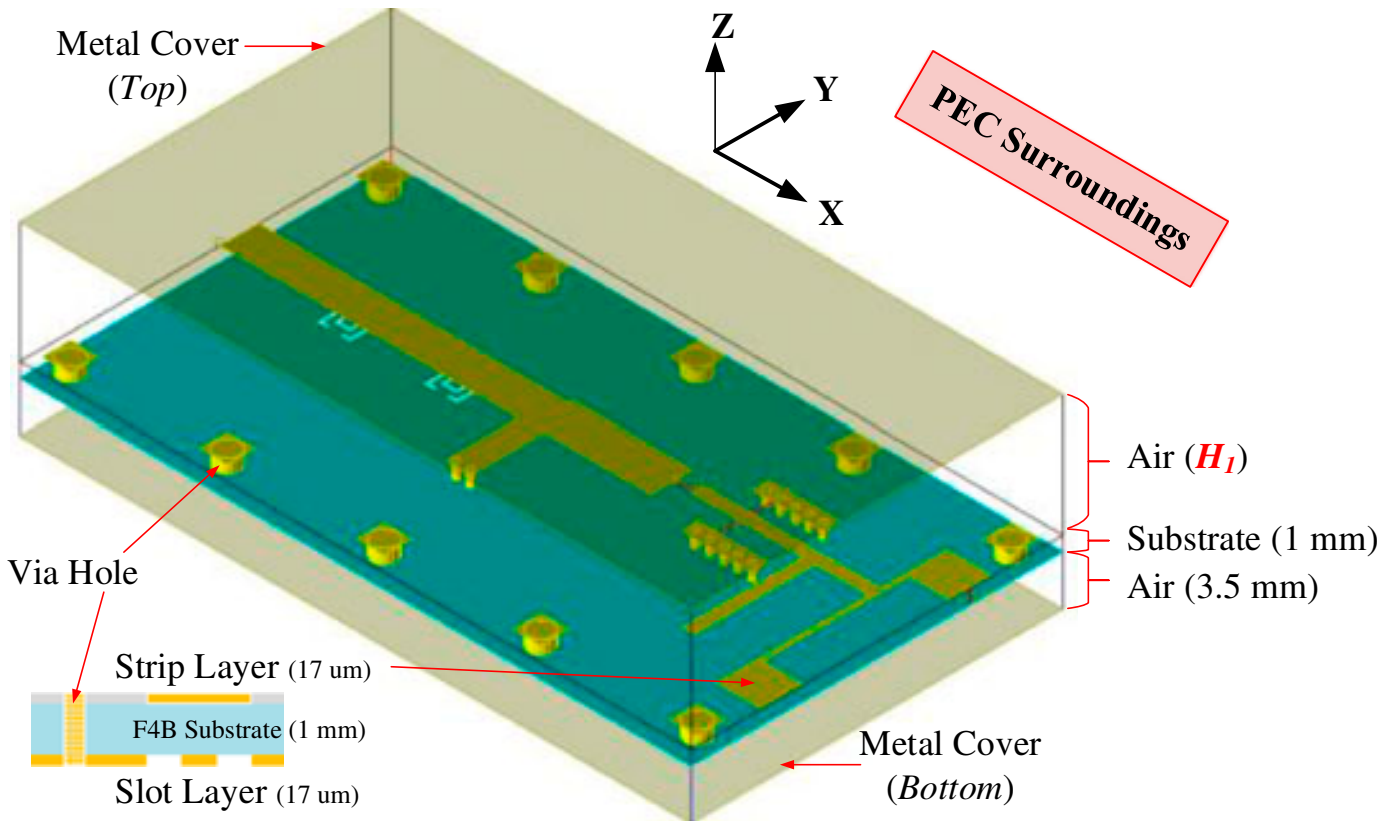

Figure 14. 3D modeling for the rectifier with cavity.

## 3. Rectenna measurement

The whole view of fabricated rectenna, composed of the aperture-coupled patch antenna array and the series diode Class-F rectifier with cavity, is shown in Figure 18. Dimensions of the antenna and rectifier are 167 mm by 160 and 54 mm by 32 mm, respectively. SMA connectors are used to provide interconnections from antenna feed to input port of the rectifier.

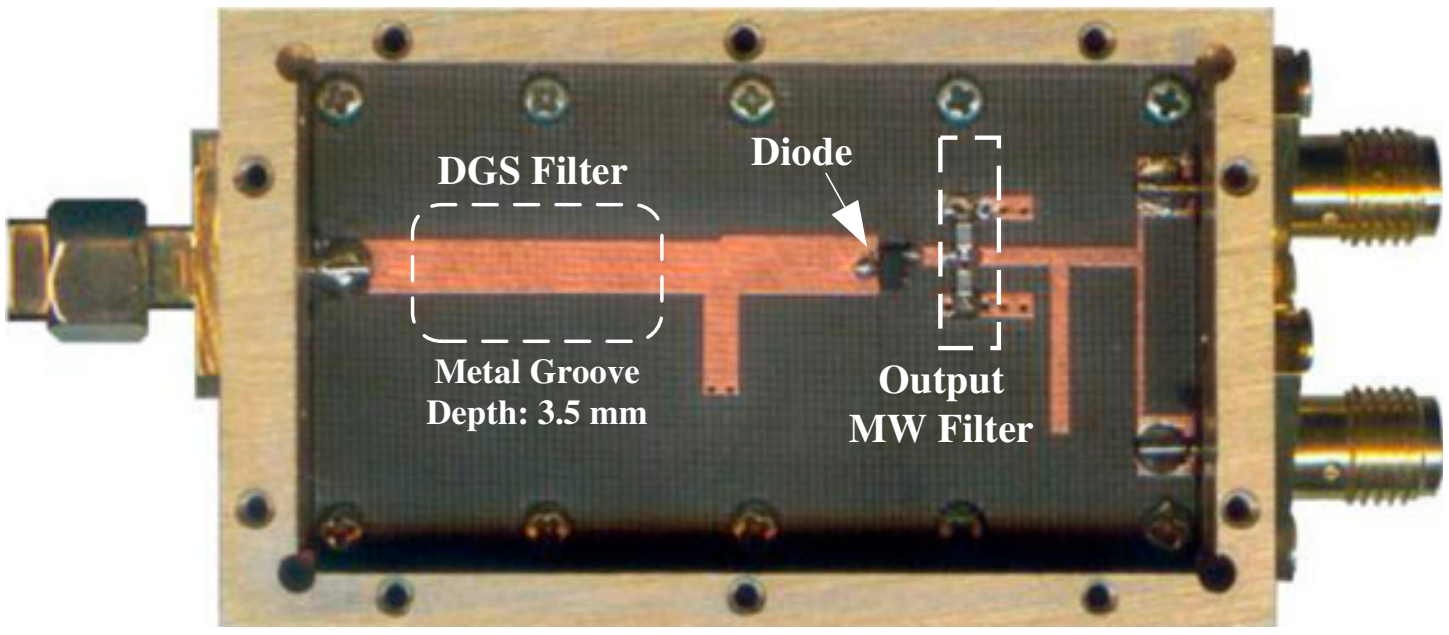

Figure 15. The fabricated 5.8 GHz Class-F rectifier with cavity.

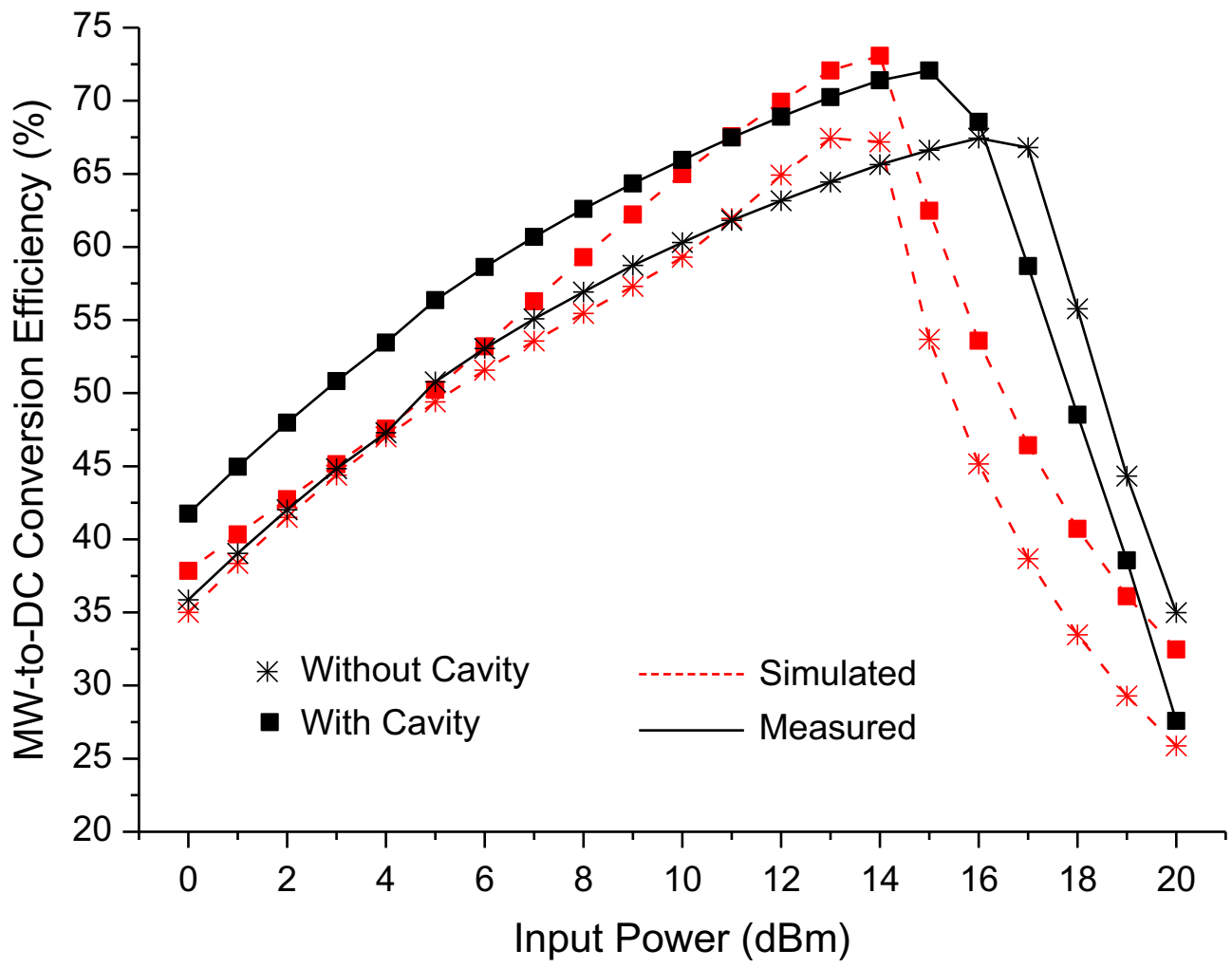

Figure 16. Efficiency comparison of the rectifier with and without cavity.

In order to measure the rectenna efficiency, a 5.8 GHz signal generated by Agilent E8267C vector signal generator is amplified and transmitted from a standard ridged horn antenna, and the output DC voltage is measured by Agilent 34970A data acquisition at the rectenna's load. The measurement setup is also a demo of MPT systems. The fixed distance of the transceiver system is 3 m ($R$) which remained in the far-field range of horn antenna. The efficiency $\eta$ can be calculated by:

$$\eta = \frac{P_{\mathrm{DC}}}{P_{\mathrm{REC}}} \tag{5}$$

where $P_{\mathrm{DC}}$ is the DC output power on the load ($R_L$) and $P_{\mathrm{REC}}$ is the received power by receiving antenna. The DC output power can be directly calculated with $V_{\mathrm{DC}}$ monitored by data acquisition and the DC load:

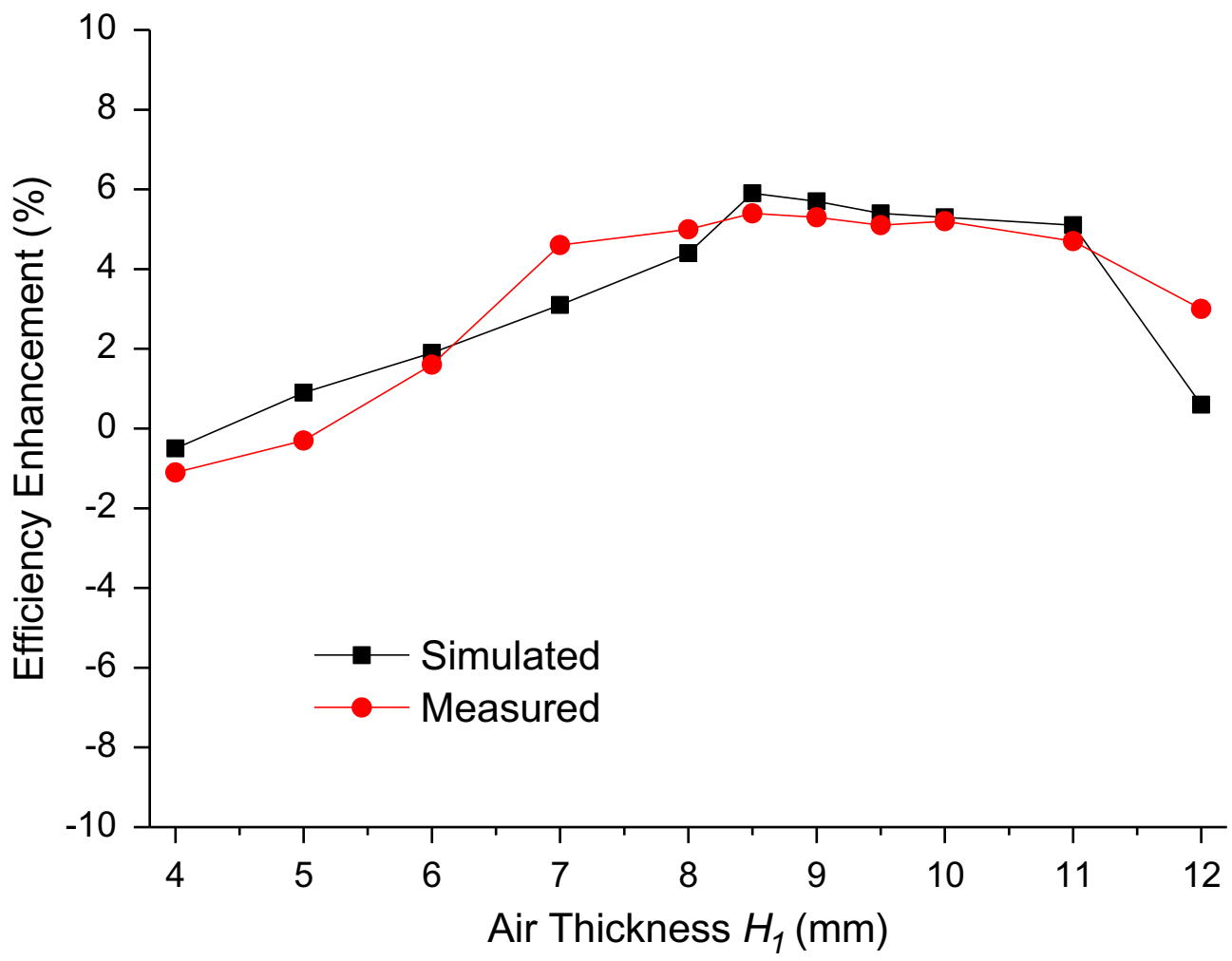

Figure 17. MW-to-DC efficiency variation relative to the air thickness $H_1$.

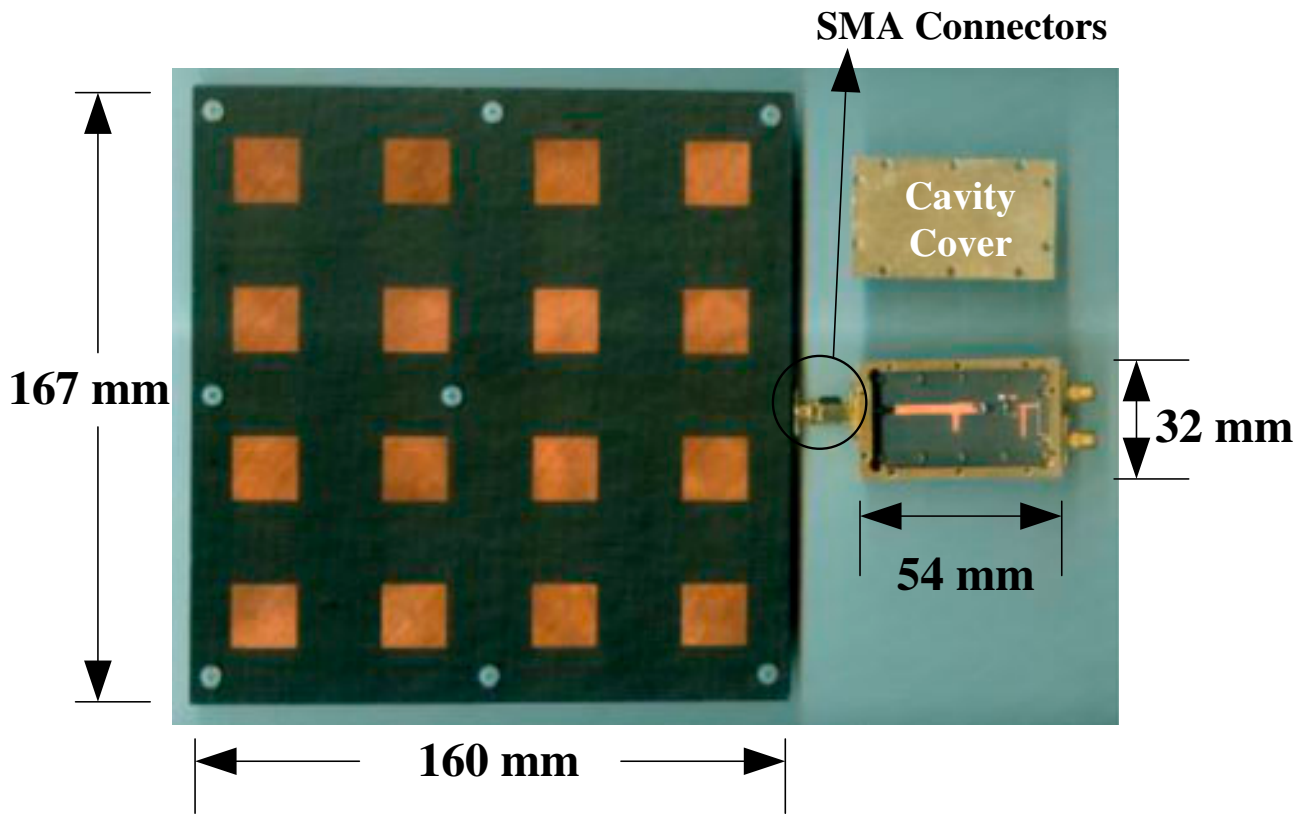

Figure 18. Photo of the proposed rectenna.

$$P_{\mathrm{DC}} = \frac{(V_{\mathrm{DC}})^2}{R_L} \tag{6}$$

The calculation process of received power is known as the Friis radio link formula, and it addresses the fundamental question of how much power is received by rectenna. As shown in Figure 19, the transmitted power ($P_T$) is monitored by a power meter at the input coupling port of coupler and calculated by Equation (7) considering the insertion loss ($IL_T$) of this coupler:

$$P_T = \frac{P_{IN}}{IL_T} \tag{7}$$

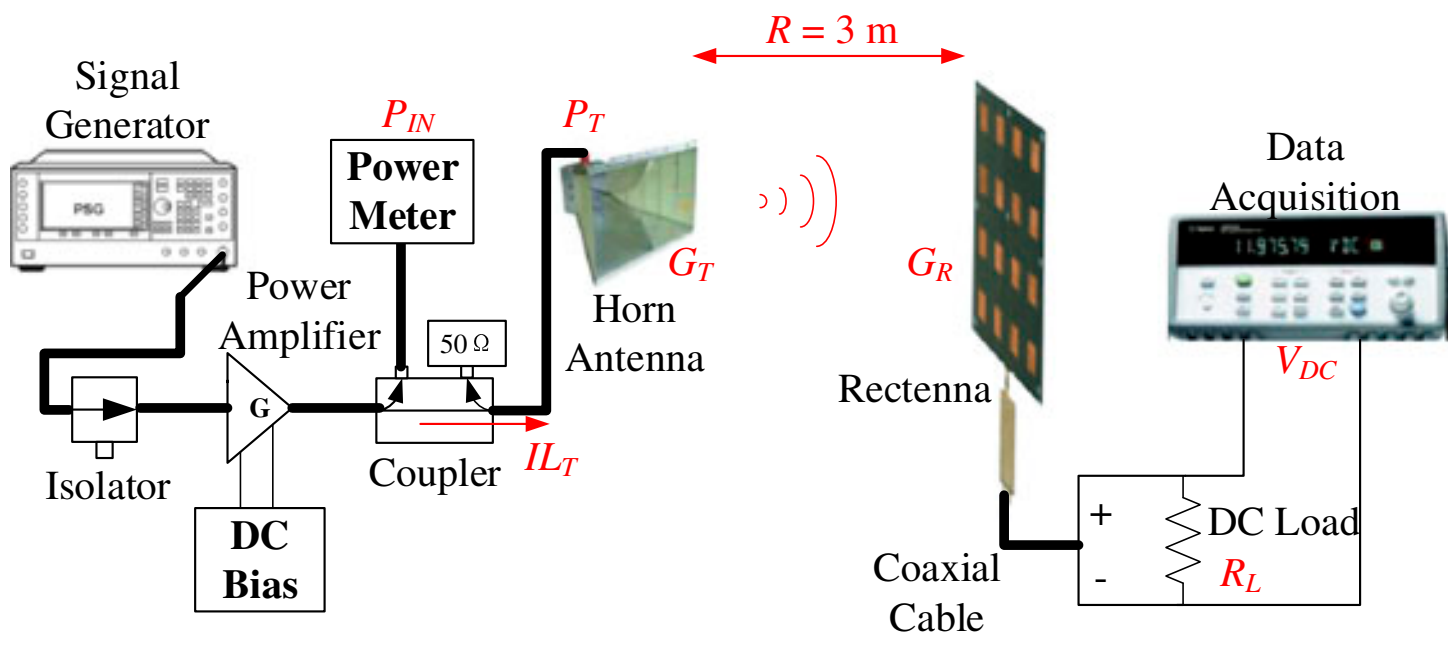

Figure 19. Equipment setup for efficiency measurement.

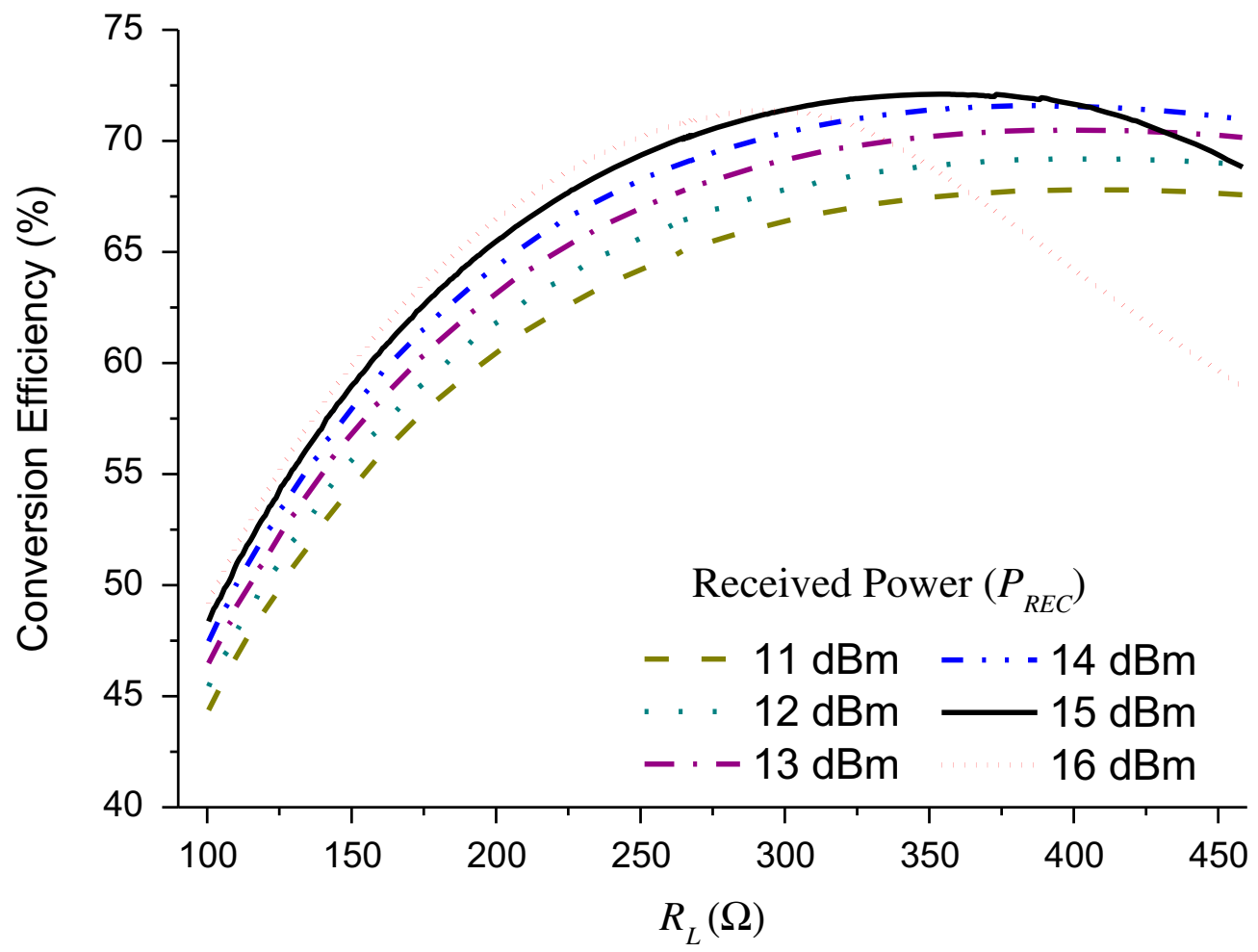

Figure 20. Measured MW-to-DC conversion efficiency for the proposed rectenna.

Then, the received power can be expressed as:

$$P_{\text{REC}} = S_{\text{AVG}} \times A_e = \frac{P_T G_T}{4\pi R^2} \times \frac{G_R \lambda_0^2}{4\pi} \tag{8}$$

where $S_{\text{AVG}}$ is the power density radiated by the horn antenna with gain ($G_T$) of 11.7 dBi, $A_e$ is effective area of rectenna with gain ($G_R$) of 19.5 dBi. $\lambda_0$ is the free space wavelength at 5.8 GHz.

In the measurement, transmitted power $P_T$ and DC load $R_L$ are varied to obtain the maximum conversion efficiency. The measured data can be seen in Figure 20. A peak efficiency of 72% was also calculated with a DC load of 380 Ω and received power of 15 dBm. The corresponding DC output voltage is approximate 3 V, nearly half of the breakdown voltage of HSMS-286.

## 4. Conclusions

In this paper, a patch antenna array based on aperture-coupled feed was developed for MPT applications with low power density. The kind of antenna feed is particularly helpful to the rectenna design due to its rejection on the coupling between its feeding circuit and radiation patch. The gain of the $4 \times 4$ antenna array reaches 19.5 dBi. The distance between feeding network layer and patch element layer has an allowable error of ±0.5 mm, which is easy to be satisfied.

A Class-F load composed of novel input and output terminations is investigated for series diode rectifier design. It is different from most conventional designs on the harmonic loads that only focus on the output circuit of a rectifier.[15] Traditional input low-pass filter [7] of a rectifier is introduced not only to block the harmonics produced by a diode, but also present desired impedance at harmonic frequencies. A whole circuit optimization for the rectifier based on a HSMS-286C diode is realized in ADS. In addition, radiation feature on the proposed rectifier is firstly analyzed and recycled with an optimized cavity. We confirm that the cavity structure leads to a 5% enhancement on MW-to-DC conversion efficiency. A MPT demo system is constructed to measure the fabricated 5.8 GHz rectenna. The MW-to-DC conversion efficiency of the proposed rectifier achieves 72%. This work is extremely valuable in the high conversion efficiency rectenna design based on harmonic termination techniques.

## Funding

This work was supported in part by the NSFC 61271074, 973 program 2013CB328902, and NCET-12-0383.

## ORCID

*Changjun Liu* http://orcid.org/0000-0003-0079-6793